# Effect of Leader's Voice on Financial Market: An Empirical Deep Learning Expedition on NASDAQ, NSE, and Beyond

Arijit Das , Tanmoy Nandi, Prasanta Saha, Suman Das, Saronyo Mukherjee, Sudip Kumar Naskar, Diganta Saha,
*Jadavpur University India*

*Abstract—* **Financial market like the price of stock, share, gold, oil, mutual funds are affected by the news and posts on social media. In this work deep learning based models are proposed to predict the trend of financial market based on NLP analysis of the twitter handles of leaders of different fields. There are many models available to predict financial market based on only the historical data of the financial component but combining historical data with news and posts of the social media like Twitter is the main objective of the present work. Substantial improvement is shown in the result. The main features of the present work are- a) proposing completely generalized algorithm which is able to generate models for any twitter handle and any financial component, b) predicting the time window for a tweet's effect on a stock price c) analyzing the effect of multiple twitter handles for predicting the trend. A detailed survey is done to find out the latest work in recent years in the similar field, find the research gap, and collect the required data for analysis and prediction. State-of-the-art algorithm is proposed and complete implementation with environment is given. An insightful trend of the result improvement considering the NLP analysis of twitter data on financial market components is shown. The Indian and USA financial markets are explored in the present work where as other markets can be taken in future. The socio-economic impact of the present work is discussed in conclusion.**

## I. INTRODUCTION

Global lifestyle and investment patterns are significantly influenced by socio-economic changes. The widespread availability of high-speed internet and online social media has enabled people to express their views freely on various scales. Nowadays, a majority of users utilize platforms such as WhatsApp, Twitter, Instagram, etc., to communicate and share comments through tweets, status updates, stories, reels, shorts, etc. The digital revolution has also transformed the financial market, making stocks, mutual funds, and precious metals easily accessible in electronic form, allowing people to make online purchases conveniently from the comfort of their homes.

Throughout the last century, economics has achieved unprecedented heights. In the multi currency system, nearly all governments worldwide strive to manage their country's economic status by boosting GDP, reducing inflation, and controlling the exchange rate between their domestic currency and foreign currency to enhance the lifestyle and income levels of their citizens. Microeconomics focuses on the financial position of entities such as farms and enterprises, guiding decision-making and choice among various options.

Stocks, mutual funds, oil, gas, precious metals, goods, services, and products witness fluctuations in their market prices over time. These variations are influenced by factors such as news, natural calamities, political stability, climate, and other dynamic elements.

The nature of news has changed considerably, with users now primarily accessing information online through various social media platforms rather than relying solely on traditional printed newspapers. Consequently, social media posts play a crucial role in shaping the financial market, especially when authored by influential leaders, significantly amplifying their impact.

In this study, we explore the impact of social media posts, such as tweets, on the financial market using artificial intelligence. We analyze the relationship between tweets and stock prices through the application of Natural Language Processing.

This motivation propels us to develop a sophisticated model that can effectively analyze the influence of social media posts, particularly tweets, on financial markets.

Novelty:

1. Beyond Traditional Models and Temporal Sentiment Dynamics: In contrast to existing financial market prediction models employed by investment banks, which predominantly rely on historical price analysis, our model extends its reach by considering the influential voices of market leaders on social media platforms. By incorporating Twitter sentiments, we introduce a state-of-the-art Memory-based Sentiment Analysis, goes beyond immediate effects by gauging the temporal impact of tweets on financial market prices.

2. Learning to Analyze Twitter Sentiment: Our model adopts a multi-step learning pro- cess, beginning with "learn to analyze twitter sentiment." This phase involves the generation of sentiment scores and categorization into positive, negative, or neutral sentiment statuses based on leaders' tweets. This state-of-the-art approach enables the system to discern the emotional tone of leaders' voices in real-time. Additionally, historical sentiment patterns from previous 'n' days to capture more enduring effects on market prices. This dual approach enhances our model's predictive accuracy by considering both immediate and longer-term impacts of social media sen- timents on financial markets.

3. Mapping Sentiments with Financial Components: Building on sentiment analysis, our model progresses

to the "learn to map tweet sentiments with different financial compo- nents". Here, the model establishes a con- nection between sentiment scores and statuses with dated market prices, effectively synchronizing the ebb and flow of sentiments with market trends. This mapping provides a comprehensive understanding of how social-media sentiments correlate with financial components.

4. Predicting Market Price Using Tweets-Stock Mapping: The final phase, "learn to predict the price using tweets-stock map- ping", equips the model with the ability to forecast financial market fluctuations based on the established mapping between social media sentiments and historical prices. This integration allows our model to offer timely and state-of-the-art predictions that consider both traditional market factors and social me- dia dynamics.

5. Back Traversal for Granular Insights: To enhance interpretability, this essentially teaches a back-traversal mechanism. By analyzing the effects of specific word or set of words on Twitter influences market fluctuations, the system gains the ability to identify and highlight the linguistic triggers influenc- ing financial markets. This granular analysis enhances the model's transparency and provides state-of-the-art insights into the relationship between social media sentiments and market movements.

**Contribution:**

1. Versatile Generalized Model: Developing a model adaptable to diverse Twitter handles and stocks without major structural changes, enhancing its generalizability.
2. Efficient Applicability: Enabling seamless application to various accounts and stocks without requiring core model modifications, streamlining the analysis process.
3. Multi-source Analysis: Utilizing tweets from multiple handles simultaneously, providing a more comprehensive view of sentiment effects on stock prices.
4. Inclusive Sentiment Consideration: Incorporating diverse sentiments from various sources, allowing the model to adapt to the richness of sentiment expressions on social media.
5. Real-time Relevance: Assigning temporal weights to tweets, prioritizing current sen- timents over older ones, mirroring the real- world scenario and enhancing the model's relevance.

We collected stock data and leaders' tweets from various exchanges and Twitter to blend finan- cial attributes with sentiments expressed in contemporary tweets using Bi-LSTM and BERT mod- els. This integration enhances our approach to predicting stock trends with a more holistic view of market sentiment.

## II.     RELATED WORKS AND LITERATURE REVIEW

The research work done in this field in the latest years are retrieved and the highlights of the important research in last five years are discussed here.

Adam Atkins et al [1] they used four datasets i.e. NASDAQ Composite, Dow Jones, Goldman Sachs and J. P. Morgan from a quantitative trading website called 'The Bonnot Gang' and used news articles from the Reuters US news archive. They used Latent Dirichlet Allocation (LDA) technique and for prediction they used the Naïve Bayes Classification Model. They showed that asset volatility can be effectively predicted by using information from news sources. They found that the model predictions were at 56.6% for DJI and 61.5% for NASDAQ.

Now, Xinyi Li et al [2] in their work they deal with NY Times editorial and social media texts and combine the stock adjusted close price. They used the VADAR to get the sentiments score then they applied the (Differential Privacy) DP-LSTM-ARIMA model. This model showed improvement over the normal LSTM time series model. The Mean Square Error for this model is 198.7500672 and the accuracy is 0.99582651.

Yinghao Ren et al [3] analyzed the news impact on stock price movements by using DBLSTM. From Sina Finance and Economics, Oriental Wealth Network, they took Chinese Financial market news data for their model. The model gives a better result than the SVM, LR models.

Xin Du et al [4] in their paper they have done portfolio optimization using Reuters & Bloomberg headlines and Wall Street Journal dataset and for stock data they used the two subsets of the S&P500 index. They proposed WA (Weighted average) +CS (classifier sharing) +DVR (dual-vector representation) model to optimize the portfolio. They got that for WSJ dataset 37% more return found and for R&B dataset 180% more return found with respect to covariance.

Sunghyuck Hong [5] in his paper he proposed deep learning-based LSTM and YTextMiner model to predict the future prices based on real time stock news and past time series analysis data. Here, Samsung Electronics past data from Yahoo finance has been used.

Isaac Kofi Nti et al [6] used historical stock price data from Ghana Stock Exchange, financial tweets posted on Twitter, Google trends to predict the movement. They used MLP-ANN model. They observed an accuracy of 51.15% based on Google trends, 57.78% based on Twitter, 41.65% based on forum post, 53.12% based on web news and 73.89% based on a combined dataset.

Sandipan Biswas et al [7] took the news articles from Yahoo Finance. They used the NLTK and VADER for their model.

Nur Ghaniaviyanto Ramadhan et al [8] took the stock dataset of Bank Mandiri of Indonesia through the yahoo finance website, and the combined this dataset with Indonesian news titles data. They made a MLP-NN. They got an accuracy of nearly 80%.

Marah-Lisanne Thormann et al [9] used the twitter data and APPLE company's stock data from Yahoo Finance and built an RNN LSTM model. This model achieved an accuracy of 87.6 % in predicting movements of the Dow Jones Industrial Average (DJIA).

Priyank Sonkiya et al [10] used a modified version of Generative Adversarial Network (GAN) and BERT to predict the stock prices using stock indexes of various countries, the technical indicators, historical prices and some commodities, along with the sentiment scores for Apple Inc. They used various stock indices data like NYSE, NASDAQ, S&P500, Indian, Hong Kong, Tokyo, and Shanghai from Yahoo Finance. The model achieved 18.2469 as RMSE for the testing data.

Zhenda Hu [11] proposed an approach that combines CEEMDAN, LSTM with attention mechanism and addition (CEEMDAN-LSTM_att-ADD) for crude oil prices. News text data from ZhongYou (http://www.cnoil.com/) and international oil (http://oil.in-en.com/) has been used. For the timeseries data daily spot prices of WTI crude oil is taken from US EIA.

Mahtab Mohtasham Khani et al [12] used tickers Gold prices from Yahoo Finance and they built some machine learning models i.e. vanilla stacked LSTMs, encoder–decoder, Bidirectional and CNN LSTM and they found that vanilla staked LSTM performed better. For two days of prediction, the model achieved $5e-4$ MSE for single-step and $8e-4$ for multi-step.

Petr Hajek et al [13] proposed a Fuzzy Unordered Rule Induction Algorithm with evolutionary tuning (FURIA + ET). They used COMEX Gold futures daily prices data for the period from 2007 to 2017 from the MarketWatch database. From the Thomson Reuters news service the news corpus was downloaded for the period (2007–2017). The test accuracy of 94.61% was obtained by the model.

Ye Ma et al [14] proposed a novel Distributed Representation of News (DRNews) model with LSTM. They used various news articles to train their model. They found that this model performs better than the BERT (Bidirectional Encoder Representations from Transformers) model in this scenario.

Taylan Kabbani et al [15] used the daily stock price data of Amazon.com Inc. (AMZN), Apple Inc. (AAPL), and Netflix Inc. (NFLX) companies from 2016-01-01 to 2020-04-01 using Yahoo Finance website and the news articles are taken from publicly available datasets. These news articles were taken from Reuters, CNN, CNBC, The New York Times, The Hill, Washington Post, and others. They proposed a VADAR + SPARK based model where VADAR is used to get the sentiments of the available news text data and the big data platform Spark is used with technical indicators like RSI, %K, SMA and some classifiers like Random Forest, Logistic Regression, and Gradient Boosting Machine. The accuracy of 0.6358 on the testing set has been achieved.

Ishu Gupta et al [16] proposed a HiSA-SMFM they mainly used the LSTM and TextBlob for their model. They Tata Motors' historical data taken from NSE (National Stock Exchange) India website, and they also collected tweets from the twitter regarding the stock. Average accuracy of 94.99% for HiSA-SMFM has been obtained.

Shayan Halder [17] proposed a deep learning based FinBERT-LSTM model which combines news sentiment and historical timeseries data to make future prediction using LSTM. He used NASDAQ-100 index stock price historical data from Yahoo Finance website to feed into the FinBERT model and he collected news articles from New York Times for the sentiments analysis. The model showed the MAE, MAPE and ACCURACY of 174.94284259, 0.01409574846 and 0.98590425153 respectively.

Zakaria Alameera et al [18] used 360 monthly observations of gold prices as the data source from "World Bank" freely available data set. They proposed the WOA-NN (Whale Optimization Neural Network) model. The model showed better results than GA–NN, PSO–NN, and GWO–NN models. The model showed RMSE, MSE, STD and $R^2$ values as 0.02131, 0.00047, 0.00340 and 0.9989 respectively.

Jessica et al [19] made a sentiment analysis+ moving average model. Here, the tweets from CNBC, Wall Street Journal, Forbes, Market Watch, and Reuters have been taken into consideration. After analysing and testing the model they found that the proposed model (MA5) + CNBC news achieved better results. The better model gave Accuracy, Precision and Recall as 0.753, 0.775 and 0.756 respectively.

Saloni Mohan et al [20] built a model by using deep learning models. They collected two different datasets for this research. The daily stock price dataset consists of closing stock prices of the S&P500 companies, from February 2013 to March 2017. They also collected news articles from February 2013 to March 2017 from international daily newspaper websites for the S&P500 companies. They found that when the RNN+ textual information is used then a better result was found. The RMSE value for this model came out as 10.43 for multivariate.

Yingzhe Dong et al [21] used tweets containing stock name keywords from some reputed Twitter accounts. The researchers proposed the BERT-LSTM (BELT) model. Here, the BERT base has been used with the LSTM deep learning model.

Ioannis E. Livieris et al [22] used daily gold prices data from Jan 2014 to Apr 2018 from Yahoo Finance Website. They proposed a CNN–LSTM. They found an accuracy of 55.26 and 51.58 for the two models.

Jingyi Shen et al [23] used 3558 Chinese stocks data. To reduce the feature space Researchers used the PCA technique. They proposed an LSTM based prediction model. The model gave binary accuracy of 0.93 and F1 score as 0.93.

Bipin Aasi et al [24] proposed a MMLSTM model to predict the Apple Inc company stock price. To train the model the $AAPL stock's historical data has been obtained from Yahoo! Finance. Google Trends data has been collected related to this company. For Apple, news headlines searched from SeekingAlpha and the tweets containing the specified keyword have been taken from Twitter. After building the model they showed that this model gives better result than the ARIMA model. Mean MAPE % found for this model was 6.328 and Mean MAAPE % was 6.311.

Wasiat Khan et al [25] proposed a model combined with social media data and financial news data for predicting stock market trends. Tweets data from twitter, stock data from New York Stock Exchange, London Stock Exchange, Microsoft Corporation, Oracle Corporation, Twitter, Inc., Motorola Solutions, Inc., Nokia Corporation etc. from Yahoo Finance, stock market related news from news websites such as Financial Times, Reuters, etc. have been collected by the researchers. Here, Sentiment analysis is performed using Stanford sentimental analysis package of Stanford NLP for the financial news and the processed tweets. Here researchers proposed a Hybrid Algorithm consisting of RF (Random Forest), ET (Extra Tree), and GBM (Gradient Boosting) classifiers. The proposed model shows an overall accuracy of 66.32%.

Naadun Sirimevan et al [26] used the DJI (Dow Jones Industrial Average Index) for this research. They got the historical stock data from Yahoo Finance, trends data from Google Trends, Twitter data by using Twitter API, Web news headlines from Reuters. They made a LSTM – RNN model and got accuracies of 0.6292, 0.6367 and 0.6702 respectively for twitter, web news and search engine query models for the 30-days prediction period.

Otabek Sattarov et al [27] built a model using a sentiment analyser on Bitcoin-related tweets and financial data. They used the VADAR technique to get the sentiment of the Tweets related to Bitcoin. From websites like BITSTAMP, COINBASE, ITBIT, KRAKEN, researchers got the daily Bitcoin historical price. They observed 62.48% accuracy when based on historical price and bitcoin-related tweet sentiment.

Padmanayana et al [28] gathered headlines from FinViz, Yahoo Finance and gathered tweets of several companies like Apple, Amazon, Microsoft using Tweepy. They used VADAR for getting the sentiment score and then they fed all data to XGBoost to predict the output. The model showed an accuracy of 89.8%.

Ashwini Saini et al [29] used the Indian Stock Market data to build the model. After comparing it with SVM, CNN etc. they found that the LSTM NN model performed better. They got the accuracy for this model as 87.86%.

Jithin Eapen et al [30] used the S&P 500 dataset from Yahoo Finance website. The researchers proposed Multiple Pipeline CNN and Bi-Directional LSTM(BILSTM) Model. They got a Mean Test Score of 0.000281317 for the 200 LSTM units.

Pengfei Yu et al [31] collected historical dataset for the Nikkei 225 (N 225), S&P 500, the Dow Jones industrial average (DJIA), the China Securities index 300 (CSI 300), the Hang Seng index (HSI) and the ChiNext index from TuShare financial data interface (tushare.org), Yahoo Finance (finance.yahoo.com) and relevant organizations. They built a phase-space reconstruction Deep neural networks long short-term memory (PSR-DNN-LSTM) model. They got the RMSE error % for S&P 500, DJIA, N 225, HIS, CSI 300, ChiNext are 7.92, 5.88, 5.60, 5.25, 5.92 and 4.15 respectively.

Md. Arif Istiake Sunny et al [32] used the Google company's historical data from Yahoo Finance website for the period of 19/08/2004 to 04/10/2019. They proposed their model using LSTM and BI-LSTM. They found that for 2 hidden layers and 50 epochs the RMSE is 0.0004219.

Sidra Mehtab et al [33] used Nifty50 index values from Yahoo Finance for the period of December 29, 2014 to July 31, 2020. They proposed CNNs, LSTM network-based predictive models. Considering previous two weeks' data as input they got the RMSE score as 0.0350 for Univariate Encoder-Decoder Conv. LSTM Model.

Adil MOGHAR et al [34] proposed a LSTM based model. They collected the historical data for the GOOGLE and NKE from Yahoo Finance website for the period from 8/19/2004 to 12/19/2019 and from 1/4/2010 to 12/19/2019 for GOOGLE and NKE respectively. After building the model the loss value of 4.97E-04 and 8.74E-04 were found for 100 epochs respectively.

Irfan Ramzan Parray et al [35] collected the NIFTY50 index of nearly all 50 stocks timeseries historical data from January 1, 2013 to December 31, 2018 using nseindia.com website. They also collected various technical indicators data like MACD, EMA, RSI and ATR. They used 3 models i.e., SVM (Support Vector Machine), perceptron neural network and logistic regression. The perceptron neural network model, SVM model and logistic regression model showed an accuracy of 76.68%, 89.93% and 89.93% respectively and F1 score as 73.61%, 89.27% and 89.87% respectively.

## III. PROPOSED METHODOLOGY

### a) Problem Statement

Given a dataset consisting of historical stock prices and a collection of tweets from prominent figures in the financial industry, the objective is to develop a predictive model that incorporates sentiment analysis of social media to enhance the accuracy of stock price forecasting. The problem can be formulated as follows:

**Given:** Historical stock price data: $\{P(t_1), P(t_2), ..., P(t_N)\}$ where $P(t_i)$ represents the stock price at time $t_i$.
Twitter sentiment data: $\{S(t_1), S(t_2), ..., S(t_N)\}$ where $S(t_i)$ represents the sentiment score derived from the tweets at time $t_i$.

**Find:** A predictive model, represented by a function f, that takes the historical stock price data and sentiment scores as inputs and predicts the future stock prices: $\{P(t_{N+1}), P(t_{N+2}), ..., P(t_{N+M})\}$ where $M > 1$.

**Objective:** Minimize the prediction error between the actual stock prices and the predicted stock prices, given the historical stock data and sentiment scores: min (P_actual - f(P(t_1), P(t_2), ..., P(t_N), S(t_1), S(t_2), ..., S(t_N))).

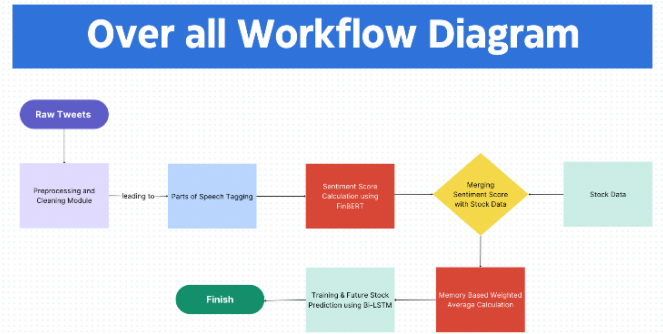

Figure 1: Workflow of the proposed model

### b) Algorithm

**Segment 1: FinBert Scoring**

/* Sentiment Analysis for Tweets */
**Initialization:** sentimentScores←zeros(N,2,3)
Input: cleanedTweets(Ш) ∈ V$^N$, models ($\theta_1$ (Prosus AI), $\theta_2$(Yiyanghkust), tokenizers ($\tau_1$, $\tau_2$).

**Output:** sentimentScores ($\omega$) $\in \mathbb{R}^{(N \times 2 \times 3)}$, where N is the number of tweets.
**Hyperparameters (hp):** $\eta$, $L$, and $H$ are hyperparameters representing the learning rate, number of layers, and number of attention heads, respectively. These values are passed to the models during training.
**Parameter:** $\theta_1$ and $\theta_2$ are parameters representing the sentiment analysis models. $\tau_1$ and $\tau_2$ are tokenizers, which are hyperparameters used for text tokenization. $\rho$ is tokenized text. P is representing pipeline( $P_1$ represents cleaned tweets in Prosus, $P_2$ represents cleaned tweets in Yiyanghkust, $P_3$ represents POS tagged tweets in Prosus, $P_4$ represents POS tagged tweets in Yiyanghkust. $clm_1$, $clm_2$, $clm_3$, $clm_4$ are representing cleaned tweets in PROSUS, cleaned tweets in Yiyanghkust, POS tagged tweets in PROSUS, POS tagged tweets in Yiyanghkust.

1. for t ∈ [Ш] :
2.     $\rho_1 \leftarrow \tau_1(t)$
3.     $\rho_2 \leftarrow \tau_2(t)$
4.     $P_1, P_2, P_3, P_4 \leftarrow \theta_1(\rho_1), \theta_2(\rho_2), \theta_1(\rho_1), \theta_2(\rho_2)$
5.     $\omega$[$clm_1$, $clm_2$, $clm_3$, $clm_4$] $\leftarrow$ [ $P_1, P_2, P_3, P_4$]
6. end
8. return $\omega$

**Segment 2: Tweet-Stock Mapping**

**Input:** tweets(T) $\in V*$, stock($S_{data}$) $\in \mathbb{R}^{d_{stocks} \times M}$, where $d_{stocks}$ is the dimension of stock data, and $M$ is the number of trading days.
**Output:** Mapped tweet-stock data $S_{mapped} \in \mathbb{R}^{d_{stocks} \times M}$.
**Hyperparameter:** '+', '-', '~' are representing 'positive', 'negative', 'neutral' sentiments.
**Parameter:** t ∈ [T], c ∈ ['+', '-', '~'] is representing tweets and sentiment labels. $\ominus$ represents scores. OneHotEncoding function is denoted by Ш. $l_{scores}$ is representing each label score. d, representing each day from 1 to M.

1. for t ∈ [T]:
2.     for c ∈ ['+', '-', '~']:
3.         $\check{e} \leftarrow$ Ш (T[t, clm])
4.         score $\leftarrow \check{e} * l_{scores}$
5.         for d ∈ D[1:M]:
6.             score[t, d] $\leftarrow \frac{\sum_{i=1}^{M} W[t, d-i] * i}{\sum_{i=1}^{M} i}$
7.         end
8.     end
9. end
10. $S_{mapped}$ ['+', '-', '~'] $\leftarrow$ score$_{1 \rightarrow M}$['+', '-', '~']
11. return $S_{mapped}$

**Segment 3: Data Scaler**

**Input:** Master Dataset(X), a dataset containing scored tweets and stock data, denoted as data $\in S_{mapped}$. $\rho$ (Train-Test Split Ratio): A ratio specifying the division between training and testing data, denoted as ratio $\in \mathbb{R}$.

**Output:** Scaled Training Data($X_þ$), the scaled training dataset, denoted as train $\in \mathbb{R}^{d_e \times l_þ}$. Scaled Testing Data($X_\sigma$), the scaled testing dataset, denoted as test $\in \mathbb{R}^{d_e \times l_\sigma}$.
Scaling Models(S), a collection of scaling models, represented as scalers $\in \mathbb{R}^{d_e \times N_{data}}$.
**Hyperparameter:** $N_{rows}$ ($N_{rows} \in \mathbb{N}$), the number of records in X. MinMaxScaler($X_\alpha$) represents the MinMaxScaler operation applied to the record $X_\alpha$. $N_þ = |\rho \times N_{rows}|$ be the number of records allocated for training, and $N_\sigma = N_{rows} - N_þ$ be the number of records allocated for testing, based on the given ratio $\rho$.
**Parameter:** $\alpha$ represents the index of the column, $\alpha \in [1: N_{rows}]$.

1. $S = \sum_{\alpha=1}^{N_{rows}} MinMaxScaler(X_\alpha)$
/*Larger section of the dataset for training */
2. $X_þ = \sum_{\alpha=1}^{N_þ} MinMaxScaler(X_\alpha)$
/* Smaller section of the dataset for testing */
3. $X_\sigma = \sum_{\alpha=N_þ+1}^{N_{rows}} MinMaxScaler(X_\alpha)$
4. return $X_þ$, $X_\sigma$, S

**Segment 4: Data Preparation**

**Initialization:** $N_þ$, total number of records in the training dataset, represented as a positive integer ($N_þ \in \mathbb{N}$). $N_\sigma$, total number of records in the test dataset, represented as a positive integer ($N_\sigma \in \mathbb{N}$).
**Input:** Train(þ), training dataset ( þ ⊆ [1, $N_þ$] ). Test($\sigma$), test dataset, ( $\sigma$ ⊆ [1, $N_\sigma$] ).
**Output:** $X_þ$, $Y_þ$, input and output sequences for training, ($X_þ$, $Y_þ$ ⊆ [1, $N_þ$]). $X_\sigma$, $Y_\sigma j$, input and output sequences for testing, ($X_\sigma$, $Y_\sigma$ ⊆[1, $N_\sigma$]).
**Hyperparameter:** Lookback(ш), number of previous records to consider, (ш ∈ $\mathbb{N}$).
**Parameter:** i and j, loop variable, (i ∈ [$N_þ$ − ш]) and (j ∈ [$N_\sigma$ − ш]).

1. for i ∈ [$N_þ$ − ш]:
2.     $X_þ \leftarrow$ þ [ i : i+ш ]
3.     $Y_þ \leftarrow$ þ [ i+ш+1 ]
4. end

5. for j ∈ [$N_\sigma$ − ш]:
6.     $X_\sigma \leftarrow \sigma$ [ j : j+ш ]
7.     $Y_\sigma \leftarrow \sigma$ [ j+ш+1 ]
8. end
9. return $X_þ$, $Y_þ$, $X_\sigma$, $Y_\sigma$

**Segment 5: Create BiLSTM Model**

**Inputs:** $X_þ \in \mathbb{R}^{[N_{samples}, ш, N_{features}]}$ /* Training set with look back */
Activation(ä), activation function name for the layers.
**Outputs:** model(M), is trained Bi-LSTM model.
**Hyperparameters:** Lookback(ш), number of previous records to consider. $\mathfrak{X}$, number of hidden units in the Bi-LSTM layers. epochs(д), number of training epochs. $B_{size}$, size of each batch during training. optimizer(Op), optimization algorithm used for training. r2$_{score}$, loss, MAE, accuracy, RMSE, evaluation metrics recorded for each epoch.
**Parameters:** X$_þ$, training set with look back. Softmax(Б), the softmax function takes a vector of raw scores (also known as logits) as input and normalizes it into a probability distribution over multiple classes.

1. М ← Seq () //sequential initialization
2. М.add(BiLSTM(Ӽ, ä, input_shape=(ɰ, N$_{features}$)))
3. М.add(BiLSTM(Ӽ, ä))
4. М.add(Dense(n, 'Б'))
5. for e ∈ [д]:
6.   М.compile(metrics=['r2$_{score}$','MAE', 'accuracy', RMSE] )
7.   М.fit(X$_þ$, д, B$_{size}$)
8. end
9. return М

### Segment 6: Fit Model

**Input:** М, represents model to be fit. X$_þ$, Y$_þ$ represents tarining data and training target data. epochs(д), number of training epochs. val, validation set split ratio. B$_{size}$, is representing batch size and patience(p) is representing early stopping callback patience.
**Output:** history(ℏ) : Fit model metrics record.
**Hyperparameter:** д, representing number of epochs. p is early stopping callback patience.
**Parameter:** М, early (e) represents an early stopping callback object and earlyStopping(e$_{stop}$) is a regularization technique used to prevent overfitting in machine learning models.

1. e ← e$_{stop}$(p)
2. ℏ ← М.fit(X$_þ$, Y$_þ$, д, val, B$_{size}$, e)
3. return ℏ

### Segment 7: Prediction

**Input:** model(М) ∈ ℝ, the trained model for prediction. X$_ϭ$ : Input sequences for testing.
**Output:** pred ∈ ℝ$^k$, predicted target (k-dimensional output).
**Parameter:** М, is trained with the given parameters.

1. pred ← М.predict(X$_ϭ$)
2. return pred

### Segment 8: Scale Inverse

**Input:** Scalers models for inverse scaling. Y$_þ$, Y$_ϭ$ , prediction - Data to be inverse scaled.
**Output:** i$_{pred}$, inverse scaled prediction. Y$_{iþ}$, inverse scaled training data. Y$_{iϭ}$, inverse scaled testing data.
**Hyperparameter:** S, Scaling models for inverse scaling.
**Parameter:** Inverse_transform(ϰ), reverses the scaling transformation, bringing the predictions back to the original scale of the data.

1. Y$_{iþ}$ ← ϰ(Y$_þ$)
2. Y$_{iϭ}$ ← ϰ(Y$_ϭ$)
3. i$_{pred}$ ← ϰ(pred)
4. return i$_{pred}$, Y$_{iþ}$, Y$_{iϭ}$

### Segment 9: Main (BiL-BERT)

**Input:** tweets(T)∈ V∗, stock(S$_{data}$) ∈ ℝ$^{d_{stocks} \times M}$, where $d_{stocks}$ is the dimension of stock data, and $M$ is the number of trading days.
**Output:** records (rec), updated records.
**Hyperparameter:** Lookback(ɰ), number of previous records to consider, (ɰ ∈ ℕ). Ӽ, number of hidden units in the Bi-LSTM layers. epochs(д), number of training epochs. B$_{size}$, size of each batch during training. optimizer(Op), optimization algorithm used for training r2$_{score}$, loss, MAE, accuracy, RMSE, evaluation metrics recorded for each epoch. Val is representing validation of split ratio. B$_{size}$, batch size, p is early stopping callback patience.
**Parameter:** М, is trained with the given parameters. activation (ä), activation function name for the layers. X$_þ$, Y$_þ$, input and output sequences for training, (X$_þ$, Y$_þ$ ⊆ [1, N$_þ$]). X$_ϭ$, Y$_{ϭj}$, input and output sequences for testing, (X$_ϭ$, Y$_ϭ$ ⊆[1, N$_ϭ$]). t ∈ [T], c ∈ ['+', '-', '~'] is representing tweets and sentiment labels. Ө represents scores. OneHotEncoding function is denoted by Ш. l$_{scores}$ is representing each label score. d, representing each day from 1 to M.

1. for s ∈ [S$_{data}$]:
2.   for hp$_{config}$ ∈ [hp]:
3.     Щ ← Tweet_cleaning(T)
4.     Ө ← FinBert_scoring(Щ)
5.     c ← name of the scored tweet
   /* column to be used for prediction ('Cleaned Tweet_PROSUS' / 'Cleaned Tweet_YIYANGHKUST' / 'POS Tagged Tweet Tweet_PROSUS' / 'POS Tagged Tweet Tweet_YIYANGHKUST') */
6.     S$_{mapped}$ ← Tweet_stock_mapping( Ө, c, S$_{data}$)
7.     þ, ϭ, S ← dataset_preprocessing(S$_{mapped}$, ρ)
8.     X$_þ$, X$_ϭ$, Y$_þ$, Y$_ϭ$ ← dataset_preparation(þ, ϭ, ɰ)
9.     М ← create_BiLSTM_model(X$_þ$, ä)
10.    ℏ ← fit_model(М, X$_þ$, Y$_þ$, д, val, B$_{size}$ , p)
11.    pred ← prediction(М, X$_ϭ$)
12.    i$_{pred}$, Y$_{iþ}$, Y$_{iϭ}$ ← scaleInverse( S, Y$_þ$, Y$_ϭ$, pred)
 /* val$_{score}$, r2$_{score}$, RMS$_{error}$, rec are calculated and listed using this model */
13.  end
14. end

c) **Experiments performed**

**Constant hyperparameters used:**

1) Twitter Handle : @narendramodi (Narendra Modi's official handle)
2) Memory factor for Memory-based Tweet-Stock Mapping Algorithm: 30 days
3) Train-Test Split Ratio: 0.8

4) Activation function used for Bi-LSTM: 'tanh'
5) Optimizer for Bi-LSTM: 'adam'
6) Patience: 15
7) Epochs: 100
8) Validation split: 0.2
9) Batch Size: 128
10) Feature Name Codes:

| Code | Feature (FinBERT Scores and Labels) |
|------|-------------------------------------|
| 1 | Cleaned Tweets with PROSUS AI Sentiment Score |
| 2 | Cleaned Tweets with Yiyanghkust Sentiment Score |
| 3 | POS Tagged Tweets with PROSUS AI Sentiment Score |
| 4 | POS Tagged Tweets with Yiyanghkust Sentiment Score |

Table 1: Feature name code

Figure 2: Yiyanghkust Word Cloud for Neutral Tweets

Figure 3: Yiyanghkust Word Cloud for Positive Tweets

Figure 4: Yiyanghkust Word Cloud for Negative Tweets

Figure 5: ProsusAI/finbert Word Cloud for Neutral Tweets

Figure 6: ProsusAI/finbert Word Cloud for Positive Tweets

Figure 7: ProsusAI/finbert Word Cloud for Negative Tweets

## IV.  DATASET AND BASELINE

1. **Dataset**

   a) **Comprehensive Dataset Compilation:** Our research work presents a comprehensive dataset encompassing historical stock prices of prominent Indian stocks (Tata Steel, NTPC, Sun Pharma, Wipro, Cipla), a leading US stock (Apple), and international indices (S&P500, Vix, Crude Oil, Hang Seng, Gold). The dataset is sourced meticulously from rep- utable exchanges, including NSE, NASDAQ, and HKSE.

   b) **Incorporation of Social Media Data:** Our research goes beyond traditional financial data by including tweets from influential personalities such as Donald Trump, Narendra Modi, Tim Cook, and prominent Twitter ac- counts like Apple News and Stocktwits. This incorporation of social media data adds a novel dimension to the analysis, capturing sentiments and opinions that can influence financial markets.
   **Tweets used for:**
   Donald Trump: S&P500, Vix, Crude Oil, Hang Seng, Gold.
   Narendra Modi: Tata Steel, NTPC, Sun Pharma, Wipro, Cipla.
   Tim Cook, Apple News, Stocktwits: Apple.

For our study, we leveraged the stock closing prices of Apple (AAPL) between March 4, 2013, and February 28, 2018, as the foundational dataset for our baseline model outlined in the paper Zhigang Jin et al [36]. Additionally, we incorporated in- formation from three distinct sources for senti- ment analysis: tweets from @applenws, which provides Apple related news on Twitter, financial discussions on StockTwits (@stocktwits), and updates from the Twitter account associated with Apple CEO Tim Cook (@timcook). This comprehensive dataset, spanning historical stock data and diverse social media sources, aims to capture both market trends and public sentiment, thereby enhancing the accuracy of our stock closing price predictions.

## 2. Baseline

A pair of crucial advancements are introduced in this study to elevate stock price prediction, build- ing upon the foundation outlined in Zhigang Jin et al [36].

Our model demonstrated superior performance in key metrics when compared to the baseline model. Specifically, it outperformed in terms of Mean Absolute Error (MAE) and Root Mean Squared Error (RMSE), indicating a higher degree of accuracy in predicting stock closing prices. Additionally, the model showcased improved accuracy and efficiency with better results in the as- sessment of time offset. These findings collectively emphasize the effectiveness of the model in surpassing the baseline, validating its predictive capabilities in financial market analysis.

## V. RESULTS AND OBSERVATIONS

### a) Experimentation Methodology

The effect of tweets by the identified stakeholders/ individuals/ organizations on the stock prices of the given scrips is observed. It represents the relationship between the tweets and the historical stock prices. A series of experiments were performed and the accurate prediction of the fluctuations of the closing prices of the stocks was observed.

**The experiments were performed per stock scrip as follows:**

1) The stock data of the stock scrip was mapped to the Tweet score data of the identified stakeholder.
2) A 30-day memory is designated for tweet-stock mapping.
3) The experiment was performed on each of the four types of FinBERT score data, namely ProsusAI FinBERT applied on both the Cleaned Tweets, and the POS Tagged Tweets, and the Yiyanghkust FinBERT-Tone applied on both the Cleaned Tweets and POS Tagged Tweets, once.
4) For each type of FinBERT score data, the look-back value was varied between 60 and 90 to predictions, and the results were observed.
5) The Validation score, R2 score, and RMSE were observed and recorded.
6) The validation and test loss plots were recorded.
7) The test-prediction plot was recorded.

The effect of the differing hyperparameters of FinBERT scoring type and look-back values are observed and recorded in tabular form. The effect the tweet sentiments have on stock prices is thus observed.

The prediction is also performed on the same scrips and indices without considering the Twitter sentiments during prediction. The raw stock data is used to train the system and collect predictions. The difference in accuracy is observed.

### b) Stock performance

The experiments focused on the NIFTY 50 stock scrips of the National Stock Exchange of In- dia and international indices, including the S&P 500 index representing the top 500 companies in the US stock markets, Chicago Board Options Exchange's CBOE Volatility Index (VIX), Gold, Crude Oil, and the Hang Seng Index of the Hong Kong Stock Exchange. we included US stocks due to their status as the world's largest economy and their developed nature, and we considered Indian stocks for their association with the world's largest population and their status as a developing coun- try. The study also considered various other coun- try exchanges, such as Hong Kong. The results showcase the performance of the top 5 stock scrips of NIFTY 50 and the 5 international indices, along with 1 US stock, in terms of prediction accuracy.

#### i. TATA STEEL

Twitter Handle used: @narendramodi (Indian Prime Minister's official Twitter handle.

| Sl No. | Feature | Look Back | Validation Score | R2 Score | RMSE |
|---|---|---|---|---|---|
| 1 | 1 | 60 | 0.669 | 0.988 | 0.037 |
| 2 | 1 | 90 | 0.630 | 0.893 | 0.049 |
| 3 | 2 | 60 | 0.616 | 0.980 | 0.044 |
| 4 | 2 | 90 | 0.630 | 0.970 | 0.037 |
| 5 | 3 | 60 | 0.634 | 0.984 | 0.038 |
| 6 | 3 | 90 | 0.651 | 0.978 | 0.039 |
| 7 | 4 | 60 | 0.627 | 0.968 | 0.043 |
| 8 | 4 | 90 | 0.620 | 0.888 | 0.040 |

Table 2: Result for Tata Steel with twitter sentiments

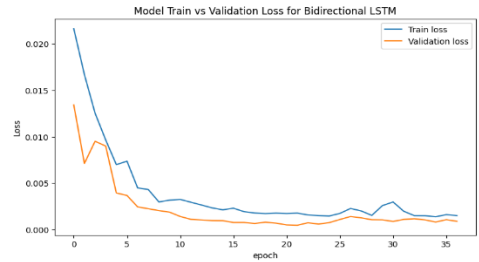

Figure 8: Training Loss and Validation Loss for Modi's Tweet & Tata Steel considering highest R2 score

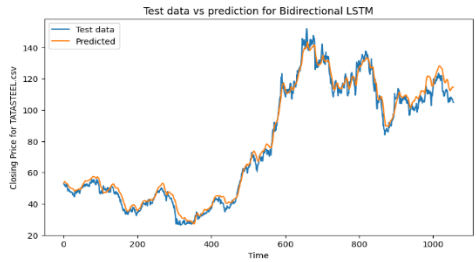

Figure 9: Actual Price and Predicted Price for Modi's Tweet & Tata Steel considering highest R2 score

| Sl No. | Look Back | Validation Score | R2 Score | RMSE |
|---|---|---|---|---|
| 1 | 60 | 0.300 | 0.987 | 0.015 |
| 2 | 90 | 0.305 | 0.934 | 0.021 |

Table 3: Result for tata steel without twitter sentiment

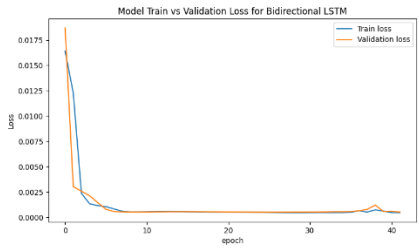

Figure 10: Training Loss and Validation Loss for Tata Steel without sentiment considering highest R2 score

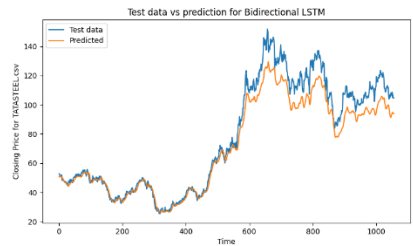

Figure 11: Actual Price and Predicted Price for Tata Steel without sentiment considering highest R2 score

ii. NTPC

Twitter Handle used: @narendramodi (Indian Prime Minister's official Twitter handle.

| Sl No. | Feature | Look Back | Validation Score | R2 Score | RMSE |
|---|---|---|---|---|---|
| 1 | 1 | 60 | 0.940 | 0.858 | 0.054 |
| 2 | 1 | 90 | 0.325 | 0.713 | 0.235 |
| 3 | 2 | 60 | 0.908 | 0.969 | 0.110 |
| 4 | 2 | 90 | 0.929 | 0.933 | 0.048 |
| 5 | 3 | 60 | 0.945 | 0.985 | 0.046 |
| 6 | 3 | 90 | 0.941 | 0.965 | 0.049 |
| 7 | 4 | 60 | 0.331 | 0.976 | 0.260 |
| 8 | 4 | 90 | 0.942 | 0.939 | 0.047 |

Table 4: Result for NTPC with twitter sentiments

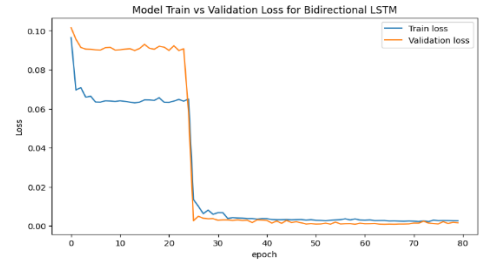

Figure 12: Training Loss and Validation Loss for Modi's Tweet & NTPC considering highest R2 score

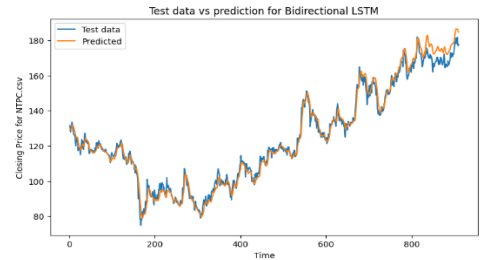

Figure 13 : Actual Price and Predicted Price for Modi's Tweet & NTPC considering highest R2 score

| Sl No. | Look Back | Validation Score | R2 Score | RMSE |
|---|---|---|---|---|
| 1 | 60 | 0.943 | 0.946 | 0.036 |
| 2 | 90 | 0.943 | 0.939 | 0.192 |

Table 5: Result for NTPC without twitter sentiment

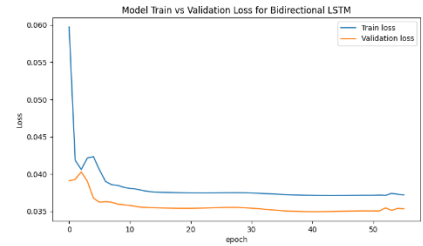

Figure 14: Training Loss and Validation Loss for NTPC without sentiment considering highest R2 score

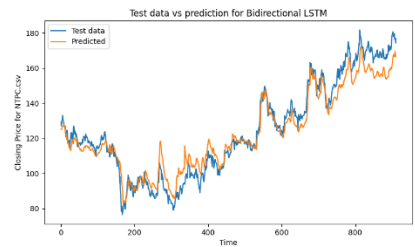

Figure 15: Actual Price and Predicted Price for NTPC without sentiment considering highest R2 score

### iii. CIPLA

Twitter Handle used: @narendramodi (Indian Prime Minister's official Twitter handle.

| Sl No. | Feature | Look Back | Validation Score | R2 Score | RMSE |
|---|---|---|---|---|---|
| 1 | 1 | 60 | 0.808 | 0.954 | 0.041 |
| 2 | 1 | 90 | 0.801 | 0.920 | 0.058 |
| 3 | 2 | 60 | 0.806 | 0.950 | 0.036 |
| 4 | 2 | 90 | 0.796 | 0.187 | 0.04 |
| 5 | 3 | 60 | 0.803 | 0.914 | 0.057 |
| 6 | 3 | 90 | 0.809 | 0.973 | 0.043 |
| 7 | 4 | 60 | 0.793 | 0.951 | 0.041 |
| 8 | 4 | 90 | 0.798 | 0.633 | 0.070 |

Table 6: Result for Cipla with twitter sentiments

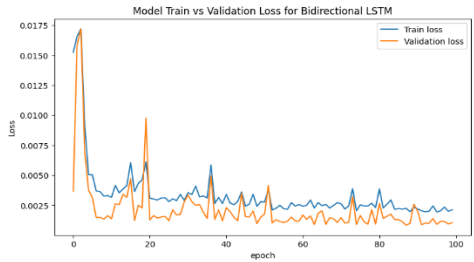

Figure 16: Training Loss and Validation Loss for Modi's Tweet & CIPLA considering highest R2 score

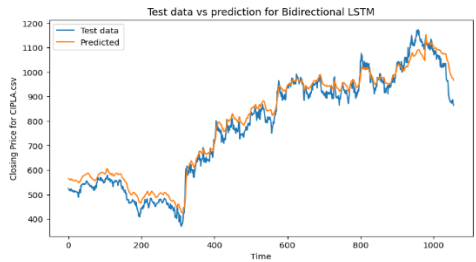

Figure 17: Actual Price and Predicted Price for Modi's Tweet & CIPLA considering highest R2 score

| Sl No. | Look Back | Validation Score | R2 Score | RMSE |
|---|---|---|---|---|
| 1 | 60 | 0.516 | 0.949 | 0.014 |
| 2 | 90 | 0.528 | 0.867 | 0.015 |

Table 7: Result for Cipla without twitter sentiment

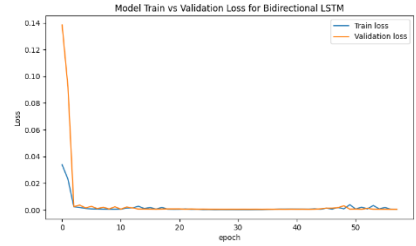

Figure 18: Training Loss and Validation Loss for CIPLA without sentiment considering highest R2 score

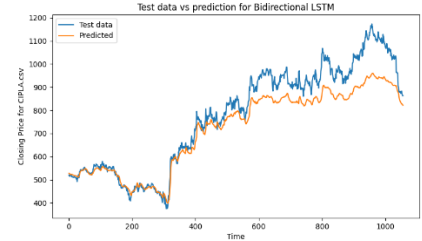

Figure 19: Actual Price and Predicted Price for CIPLA without sentiment considering highest R2 score

### iv. SUNPHARMA

Twitter Handle used: @narendramodi (Indian Prime Minister's official Twitter handle.

| Sl No. | Feature | Look Back | Validation Score | R2 Score | RMSE |
|---|---|---|---|---|---|
| 1 | 1 | 60 | 0.690 | 0.972 | 0.047 |
| 2 | 1 | 90 | 0.813 | 0.919 | 0.053 |
| 3 | 2 | 60 | 0.825 | 0.970 | 0.045 |
| 4 | 2 | 90 | 0.805 | 0.87 | 0.055 |
| 5 | 3 | 60 | 0.710 | 0.951 | 0.049 |
| 6 | 3 | 90 | 0.845 | 0.934 | 0.061 |
| 7 | 4 | 60 | 0.752 | 0.949 | 0.072 |
| 8 | 4 | 90 | 0.739 | 0.657 | 0.058 |

Table 8: Result for Sunpharma with twitter sentiments

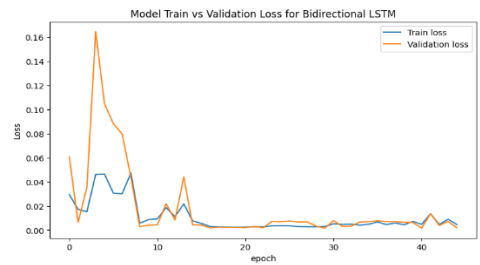

Figure 20: Training Loss and Validation Loss for Modi's Tweet & SUNPHARMA considering highest R2 score

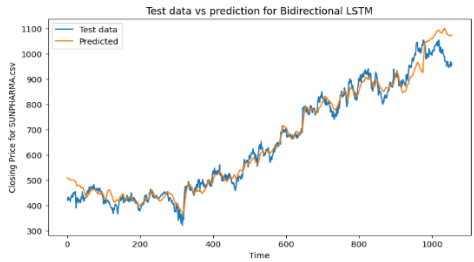

Figure 21: Actual Price and Predicted Price for Modi's Tweet & SUNPHARMA considering highest R2 score

| Sl No. | Look Back | Validation Score | R2 Score | RMSE |
|---|---|---|---|---|
| 1 | 60 | 0.463 | 0.936 | 0.035 |
| 2 | 90 | 0.652 | 0.965 | 0.030 |

Table 9: Result for Sunpharma without twitter sentiment

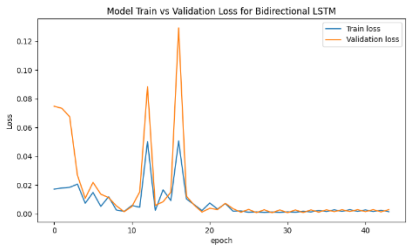

Figure 22: Training Loss and Validation Loss for SUNPHARMA without sentiment considering highest R2 score

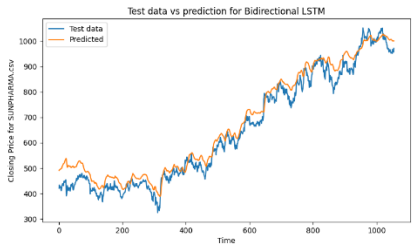

Figure 23: Actual Price and Predicted Price for SUNPHARMA without sentiment considering highest R2 score

v.     WIPRO

Twitter Handle used: @narendramodi (Indian Prime Minister's official Twitter handle.

| Sl No. | Feature | Look Back | Validation Score | R2 Score | RMSE |
|---|---|---|---|---|---|
| 1 | 1 | 60 | 0.672 | 0.917 | 0.052 |
| 2 | 1 | 90 | 0.675 | 0.878 | 0.037 |
| 3 | 2 | 60 | 0.657 | 0.800 | 0.034 |
| 4 | 2 | 90 | 0.664 | 0.802 | 0.033 |
| 5 | 3 | 60 | 0.680 | 0.857 | 0.049 |
| 6 | 3 | 90 | 0.681 | 0.549 | 0.049 |
| 7 | 4 | 60 | 0.669 | 0.896 | 0.043 |
| 8 | 4 | 90 | 0.674 | 0.960 | 0.034 |

Table 10: Result for Wipro with twitter sentiments

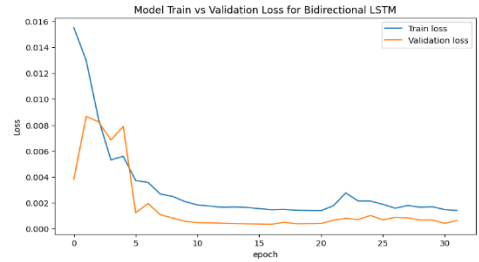

Figure 24: Training Loss and Validation Loss for Modi's Tweet & WIPRO considering highest R2 score

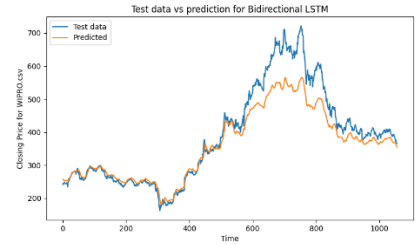

Figure 25: Actual Price and Predicted Price for Modi's Tweet & WIPRO considering highest R2 score

| Sl No. | Look Back | Validation Score | R2 Score | RMSE |
|---|---|---|---|---|
| 1 | 60 | 0.379 | 0.613 | 0.009 |
| 2 | 90 | 0.399 | 0.874 | 0.007 |

Table 11: Result for Wipro without twitter sentiment

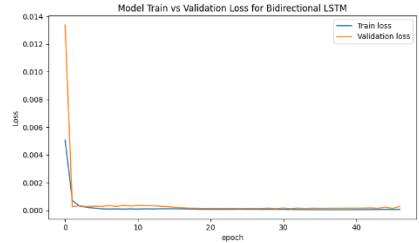

Figure 26: Training Loss and Validation Loss for WIPRO without sentiment considering highest R2 score

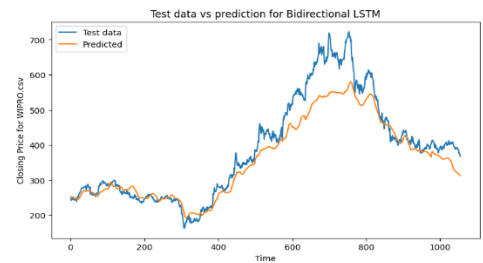

Figure 27: Actual Price and Predicted Price for WIPRO without sentiment considering highest R2 score

### vi. S&P500

Twitter Handle used: @realDonaldTrump (American previous president's Twitter handle).

| Sl No. | Feature | Look Back | Validation Score | R2 Score | RMSE |
|---|---|---|---|---|---|
| 1 | 1 | 60 | 0.881 | 0.901 | 0.044 |
| 2 | 1 | 90 | 0.881 | 0.909 | 0.067 |
| 3 | 2 | 60 | 0.853 | 0.653 | 0.055 |
| 4 | 2 | 90 | 0.888 | 0.798 | 0.798 |
| 5 | 3 | 60 | 0.891 | 0.264 | 0.264 |
| 6 | 3 | 90 | 0.891 | 0.872 | 0.872 |
| 7 | 4 | 60 | 0.881 | 0.888 | 0.888 |
| 8 | 3 | 90 | 0.887 | 0.568 | 0.055 |
| 9 | 1 | 5 | 0.868 | 0.969 | 0.066 |
| 10 | 1 | 10 | 0.858 | 0.909 | 0.072 |
| 11 | 1 | 20 | 0.867 | 0.918 | 0.066 |
| 12 | 1 | 30 | 0.872 | 0.874 | 0.038 |
| 13 | 2 | 5 | 0.885 | 0.807 | 0.037 |
| 14 | 2 | 10 | 0.876 | 0.790 | 0.058 |
| 15 | 2 | 30 | 0.890 | 0.835 | 0.037 |
| 16 | 3 | 5 | 0.878 | 0.940 | 0.037 |
| 17 | 3 | 10 | 0.883 | 0.918 | 0.060 |
| 18 | 4 | 20 | 0.884 | 0.933 | 0.059 |
| 19 | 3 | 30 | 0.882 | 0.738 | 0.061 |
| 20 | 4 | 5 | 0.875 | 0.735 | 0.036 |
| 21 | 4 | 10 | 0.868 | 0.505 | 0.056 |
| 22 | 4 | 20 | 0.873 | 0.910 | 0.054 |
| 23 | 4 | 30 | 0.868 | 0.548 | 0.077 |

Table 12: Result for S&P500 with twitter sentiments

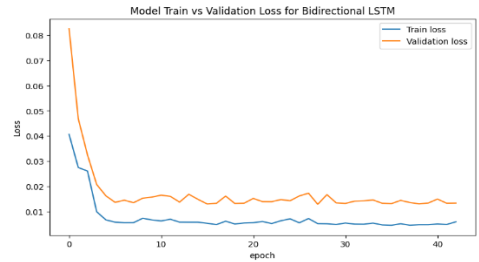

Figure 28: Training Loss and Validation Loss for Trump's Tweet & S&P500 considering highest R2 score

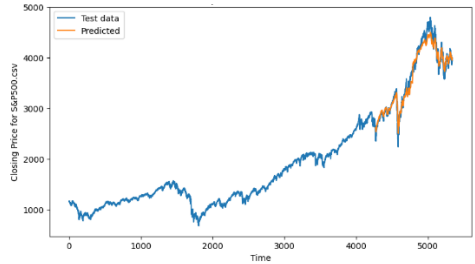

Figure 29: Actual Price and Predicted Price for Trump's Tweet & S&P500 considering highest R2 score

| Sl No. | Look Back | Validation Score | R2 Score | RMSE |
|---|---|---|---|---|
| 1 | 5 | 0.823 | 0.923 | 0.0277 |
| 2 | 20 | 0.806 | 0.964 | 0.025 |
| 3 | 60 | 0.806 | 0.772 | 0.026 |
| 4 | 90 | 0.817 | 0.778 | 0.024 |

Table 13: Result for S&P500 without twitter sentiment

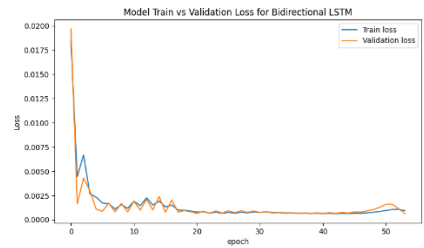

Figure 30: Training Loss and Validation Loss for S&P500 without sentiment considering highest R2 score

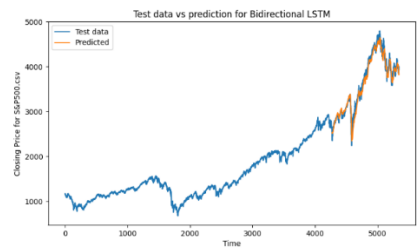

Figure 31: Actual Price and Predicted Price for S&P500 without sentiment considering highest R2 score

### vii. VIX

Twitter Handle used: @realDonaldTrump (American previous president's Twitter handle).

| Sl No. | Feature | Look Back | Validation Score | R2 Score | RMSE |
|---|---|---|---|---|---|
| 1 | 1 | 60 | 0.916 | 0.950 | 0.030 |
| 2 | 1 | 90 | 0.914 | 0.858 | 0.064 |
| 3 | 2 | 60 | 0.924 | 0.936 | 0.047 |
| 4 | 2 | 90 | 0.924 | 0.955 | 0.030 |
| 5 | 3 | 60 | 0.923 | 0.926 | 0.033 |
| 6 | 3 | 90 | 0.924 | 0.851 | 0.029 |
| 7 | 3 | 60 | 0.921 | 0.916 | 0.050 |
| 8 | 3 | 90 | 0.917 | 0.872 | 0.036 |
| 9 | 1 | 10 | 0.917 | 0.856 | 0.029 |
| 10 | 1 | 20 | 0.914 | 0.949 | 0.063 |
| 11 | 1 | 30 | 0.916 | 0.908 | 0.062 |
| 12 | 2 | 5 | 0.921 | 0.925 | 0.034 |
| 13 | 2 | 10 | 0.923 | 0.897 | 0.033 |
| 14 | 2 | 20 | 0.924 | 0.881 | 0.047 |
| 15 | 2 | 30 | 0.921 | 0.934 | 0.047 |
| 16 | 3 | 5 | 0.922 | 0.905 | 0.035 |
| 17 | 3 | 10 | 0.924 | 0.948 | 0.030 |
| 18 | 3 | 20 | 0.926 | 0.913 | 0.056 |
| 19 | 3 | 30 | 0.924 | 0.856 | 0.031 |
| 20 | 4 | 5 | 0.922 | 0.927 | 0.033 |
| 21 | 4 | 10 | 0.920 | 0.635 | 0.049 |
| 22 | 4 | 20 | 0.920 | 0.918 | 0.034 |
| 23 | 4 | 30 | 0.923 | 0.945 | 0.045 |

Table 14: Result for VIX with twitter sentiments

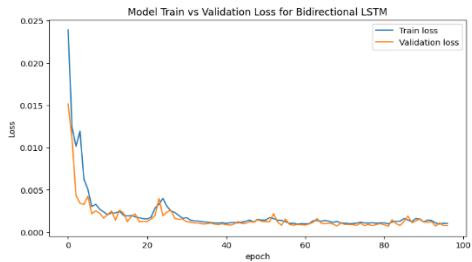

Figure 32: Training Loss and Validation Loss for Trump's Tweet & VIX considering validation score & highest R2 score

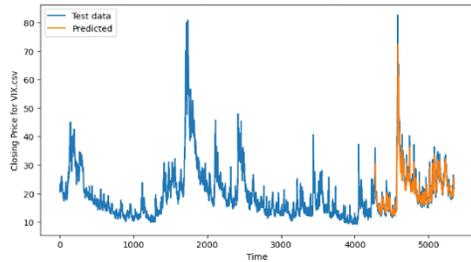

Figure 33 : Actual Price and Predicted Price for Trump's Tweet & VIX considering validation score & highest R2 score

| Sl No. | Look Back | Validation Score | R2 Score | RMSE |
|---|---|---|---|---|
| 1 | 3 | 0.871 | 0.903 | 0.018 |
| 2 | 60 | 0.869 | 0.921 | 0.016 |
| 3 | 90 | 0.866 | 0.914 | 0.017 |

Table 15: Result for VIX without twitter sentiment

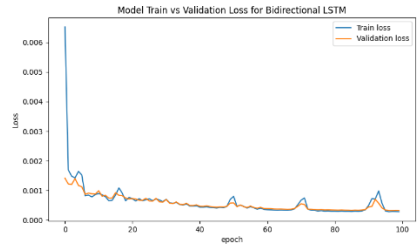

Figure 34: Training Loss and Validation Loss for VIX without sentiment considering highest R2 score

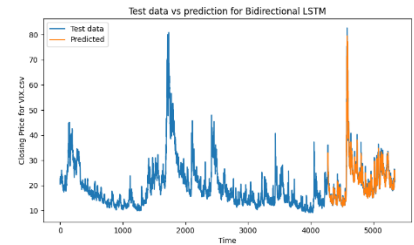

Figure 35: Actual Price and Predicted Price for VIX without sentiment considering highest R2 score

### viii. CRUDE OIL

Twitter Handle used: @realDonaldTrump (American previous president's Twitter handle).

| Sl No. | Feature | Look Back | Validation Score | R2 Score | RMSE |
|---|---|---|---|---|---|
| 1 | 1 | 60 | 0.813 | 0.935 | 0.034 |
| 2 | 1 | 90 | 0.662 | 0.797 | 0.193 |
| 3 | 2 | 60 | 0.757 | 0.971 | 0.028 |
| 4 | 2 | 90 | 0.616 | 0.965 | 0.295 |
| 5 | 3 | 60 | 0.839 | 0.928 | 0.051 |
| 6 | 3 | 90 | 0.825 | 0.911 | 0.061 |
| 7 | 4 | 60 | 0.602 | 0.316 | 0.201 |
| 8 | 4 | 90 | 0.864 | 0.972 | 0.053 |
| 9 | 1 | 5 | 0.861 | 0.963 | 0.094 |
| 10 | 1 | 10 | 0.687 | 0.921 | 0.189 |
| 11 | 1 | 20 | 0.726 | 0.386 | 0.181 |
| 12 | 1 | 30 | 0.781 | 0.905 | 0.066 |
| 13 | 2 | 5 | 0.863 | 0.748 | 0.072 |
| 14 | 2 | 10 | 0.899 | 0.958 | 0.047 |
| 15 | 2 | 20 | 0.868 | 0.953 | 0.038 |
| 16 | 2 | 30 | 0.498 | 0.830 | 0.263 |
| 17 | 3 | 5 | 0.639 | 0.931 | 0.196 |
| 18 | 3 | 10 | 0.863 | 0.664 | 0.083 |
| 19 | 3 | 20 | 0.860 | 0.964 | 0.068 |
| 20 | 3 | 30 | 0.635 | 0.799 | 0.190 |
| 21 | 4 | 5 | 0.902 | 0.972 | 0.046 |
| 22 | 4 | 10 | 0.602 | 0.848 | 0.199 |
| 23 | 4 | 20 | 0.583 | 0.894 | 0.193 |

| | 24 | 4 | 30 | 0.879 | 0.948 | 0.034 |

Table 16: Result for Crude Oil with twitter sentiments

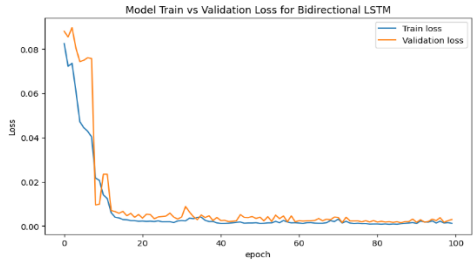

Figure 36: Training Loss and Validation Loss for Trump's Tweet & CRUDE OIL considering highest R2 score & minimum RMSE score

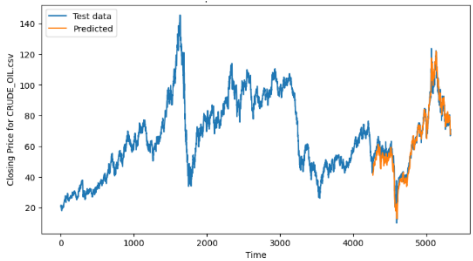

Figure 37: Actual Price and Predicted Price for Trump's Tweet & CRUDE OIL considering highest R2 score & minimum RMSE score

| Sl No. | Look Back | Validation Score | R2 Score | RMSE |
|---|---|---|---|---|
| 1 | 60 | 0.553 | 0.979 | 0.022 |
| 2 | 90 | 0.441 | 0.965 | 0.219 |

Table 17: Result for Crude Oil without twitter sentiment

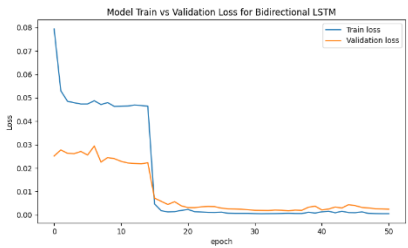

Figure 38: Training Loss and Validation Loss for CRUDE OIL without sentiment considering highest R2 score

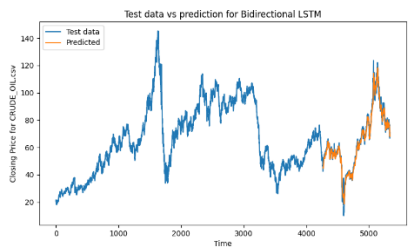

Figure 39: Actual Price and Predicted Price for CRUDE OIL without sentiment considering highest R2 score

### ix. HANG SENG

Twitter Handle used: @realDonaldTrump (American previous president's Twitter handle).

| Sl No. | Feature | Look Back | Validation Score | R2 Score | RMSE |
|---|---|---|---|---|---|
| 1 | 1 | 60 | 0.662 | 0.930 | 0.044 |
| 2 | 1 | 90 | 0.666 | 0.857 | 0.070 |
| 3 | 2 | 60 | 0.241 | 0.945 | 0.205 |
| 4 | 2 | 90 | 0.733 | 0.949 | 0.050 |
| 5 | 3 | 60 | 0.717 | 0.937 | 0.062 |
| 6 | 3 | 90 | 0.704 | 0.745 | 0.069 |
| 7 | 4 | 60 | 0.730 | 0.932 | 0.036 |
| 8 | 4 | 90 | 0.733 | 0.749 | 0.051 |
| 9 | 1 | 5 | 0.664 | 0.921 | 0.089 |
| 10 | 1 | 10 | 0.630 | 0.782 | 0.041 |
| 11 | 1 | 20 | 0.672 | 0.858 | 0.067 |
| 12 | 1 | 30 | 0.668 | 0.896 | 0.036 |
| 13 | 2 | 5 | 0.682 | 0.897 | 0.076 |
| 14 | 2 | 10 | 0.696 | 0.882 | 0.052 |
| 15 | 2 | 20 | 0.729 | 0.699 | 0.035 |
| 16 | 2 | 30 | 0.717 | 0.921 | 0.049 |
| 17 | 3 | 5 | 0.728 | 0.495 | 0.035 |
| 18 | 3 | 10 | 0.272 | 0.675 | 0.191 |
| 19 | 3 | 20 | 0.711 | 0.912 | 0.060 |
| 20 | 3 | 30 | 0.237 | 0.590 | 0.191 |
| 21 | 4 | 5 | 0.267 | 0.747 | 0.200 |
| 22 | 4 | 10 | 0.732 | 0.871 | 0.049 |
| 23 | 4 | 20 | 0.714 | 0.920 | 0.049 |
| 24 | 4 | 30 | 0.727 | 0.856 | 0.055 |

Table 18: Result for Hang Seng with twitter sentiments

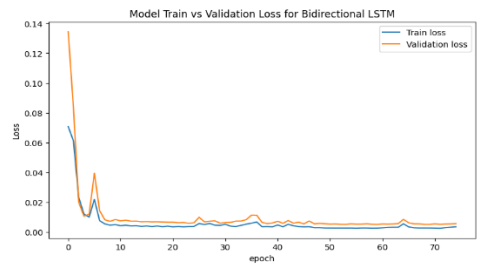

Figure 40: Training Loss and Validation Loss for Trump's Tweet & HANG SENG considering R2 score, Validation Score and RMSE score

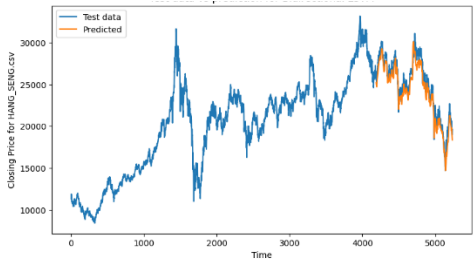

Figure 41: Actual Price and Predicted Price for Trump's Tweet & HANG SENG considering R2 score, Validation Score and RMSE score

| Sl No. | Look Back | Validation Score | R2 Score | RMSE |
|---|---|---|---|---|
| 1 | 30 | 0.444 | 0.967 | 0.029 |
| 2 | 60 | 0.435 | 0.973 | 0.030 |
| 3 | 90 | 0.426 | 0.960 | 0.034 |

Table 19: Result for hang seng without twitter sentiment

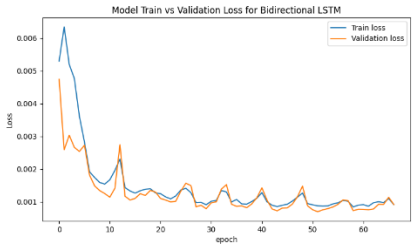

Figure 42: Training Loss and Validation Loss for HANG SENG without sentiment considering highest R2 score

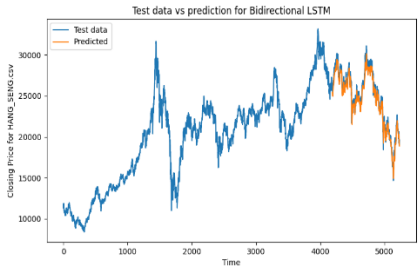

Figure 43: Actual Price and Predicted Price for HANG SENG without sentiment considering highest R2 score

x. GOLD

Twitter Handle used: @realDonaldTrump (American previous president's Twitter handle).

| Sl No. | Feature | Look Back | Validation Score | R2 Score | RMSE |
|---|---|---|---|---|---|
| 1 | 1 | 5 | 0.877 | 0.625 | 0.098 |
| 2 | 2 | 4 | 0.923 | 0.638 | 0.044 |
| 3 | 2 | 5 | 0.921 | 0.623 | 0.178 |
| 4 | 2 | 6 | 0.929 | 0.635 | 0.073 |
| 5 | 3 | 4 | 0.912 | 0.610 | 0.195 |
| 6 | 4 | 5 | 0.929 | 0.667 | 0.100 |
| 7 | 4 | 7 | 0.936 | 0.639 | 0.046 |

Table 20: Result for gold with twitter sentiments

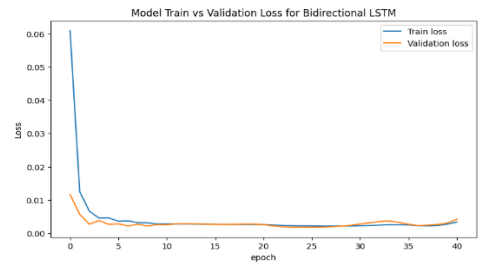

Figure 44: Training Loss and Validation Loss for Trump's Tweet & GOLD considering highest Validation score

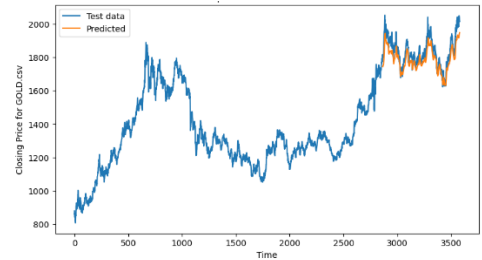

Figure 45: Actual Price and Predicted Price for Trump's Tweet & GOLD considering highest Validation score

| Sl No. | Feature | Look Back | Validation Score | R2 Score | RMSE |
|---|---|---|---|---|---|
| 1 | 3 | 90 | 0.978 | 0.162 | 0.031 |
| 2 | 3 | 60 | 0.979 | -0.329 | 0.031 |
| 3 | 3 | 30 | 0.969 | -0.738 | 0.035 |
| 4 | 3 | 10 | 0.960 | 0.385 | 0.038 |
| 5 | 4 | 20 | 0.964 | 0.592 | 0.042 |
| 6 | 4 | 5 | 0.939 | 0.242 | 0.272 |
| 7 | 4 | 10 | 0.961 | -0.880 | 0.098 |
| 8 | 4 | 60 | 0.992 | 0.021 | 0.038 |
| 9 | 4 | 90 | 0.995 | 0.517 | 0.060 |
| 10 | 4 | 30 | 0.988 | 0.694 | 0.039 |

Table 21: Result for gold without twitter sentiment

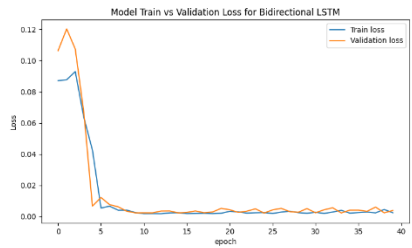

Figure 46: Training Loss and Validation Loss for GOLD without sentiment considering highest R2 score

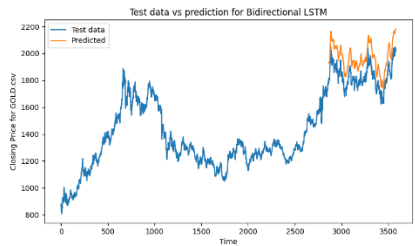

Figure 47: Actual Price and Predicted Price for GOLD without sentiment considering highest R2 score

xi.  Apple

Twitter Handle used: @timcook, @stocktwits and @applenws.

| Sl No. | Look Back | Validation Score | R2 Score | RMSE |
|---|---|---|---|---|
| 1 | 4 | 0.351 | -0.207 | 0.199 |
| 2 | 15 | 0.330 | -0.141 | 0.043 |
| 3 | 30 | 0.362 | -3.115 | 0.042 |
| 4 | 60 | 0.355 | 0.309 | 0.039 |
| 5 | 90 | 0.298 | 0.844 | 0.038 |
| 6 | 180 | 0.357 | -1.497 | 0.041 |

Table 22: Result for Apple with twitter sentiments

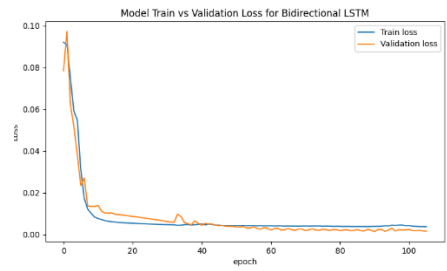

Figure 48: Training Loss and Validation Loss for Apple with sentiment considering highest R2 score

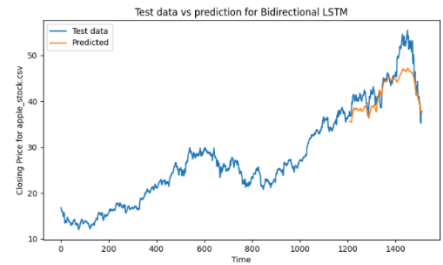

Figure 49: Actual Price and Predicted Price for Apple with sentiment considering highest R2 score

| Sl No. | Look Back | Validation Score | R2 Score | RMSE |
|---|---|---|---|---|
| 1 | 30 | 0.631 | 0.073 | 0.032 |
| 2 | 60 | 0.459 | 0.685 | 0.119 |

Table 23: Result for Apple without twitter sentiments

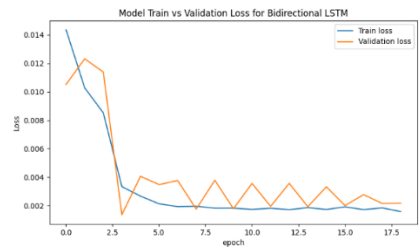

Figure 50: Training Loss and Validation Loss for Apple without sentiment considering highest R2 score

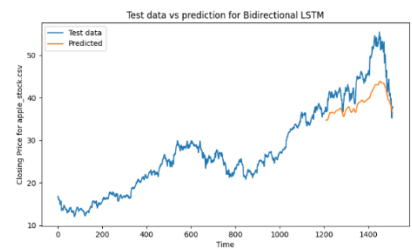

Figure 51: Actual Price and Predicted Price for Apple without sentiment considering highest R2 score

c) SOME CLOSE OBSERVATION

The model is tested on top 50 companies of NIFTY India, five companies of NASDAQ, Standard and Poor 500 of USA, and commodities like crude oil, gold etc. Here due to space constraint, top companies and commodities from different sectors are presented. To measure the effect of tweets or news on shares and stocks price, handles of leaders of both countries having highest tweeter followers, their official handles like PMO India and Presidents Office of USA, business leaders, concerned Managing Director and CEO of the companies are considered. To examine the effect of tweets over time, a large time window (look back) is taken. By analyzing the results following inferences are drawn-

1. The USA share market gives the best prediction on past data of average 60 days without considering the sentiment effect of tweets.

2. The Indian share market gives the best prediction on past data of average 75 days without considering the sentiment effect of tweets.

3. The volatility of USA share market is more than Indian share market on average 1.5 times.

4. The effect of tweets is generally dried out after three months or ninety days.

5. In case of commodities like gold, crude oil look-back period is short. For Gold the effect is maximum in the window of 4-5 days and for crude oil the effect persists maximum up to sixty days or two months.

6. Yiyanghkust model for Financial Sentiment Analysis performs better than PROSUS model and the difference is more prominent in the analysis of tweets originated from USA financial sentiment score calculation and prediction purpose.

7. POS tagged tweets give better result than untagged tweets for financial sentiment score calculation and prediction purpose.

8. The effect of tweet is lesser in the old and blue-chip companies than new and midcap companies.

9. Tweets originated from India has more effect on domestic market than tweets originated from USA.

10. Effect of tweet is more but short-lived for gold than stocks or other commodities.

## VI. EXPERIMENTAL RESULTS AND PERFORMANCE COMPARISON

In the subsequent evaluation, predictions of stock closing prices with a time offset (t) were assessed, and the accuracy (ACC) was measured under the corresponding time offset. Detailed results, in- cluding metrics such as MAE, R2, RMSE, and time delay, were presented in Table 24. The introduction of sentiment orientation demonstrated notable improvements in prediction accuracy over the baseline, showcasing enhanced RMSE and R2 results. Similarly, the attention-enhanced LSTM model exhibited effectiveness in stock closing price prediction, achieving superior results when compared to the baseline. Importantly, the time delay for predicting stock closing prices based on 60-day historical data significantly reduced from 9 to 1 day, emphasizing a substantial enhancement in model responsiveness. These findings collectively highlight the effectiveness of the proposed enhancements in refining and advancing the baseline model for stock closing price prediction.

| Model | MAE | R2 | RMSE | T | ACC |
|---|---|---|---|---|---|
| LSTM | 7.032 | 0.832 | 8.712 | 9 | 0.601 |
| LS_RF | 4.713 | 0.927 | 5.756 | 7 | 0.635 |
| S_LSTM | 3.32 | 0.956 | 4.483 | 5 | 0.657 |
| S_AM_LSTM | 2.649 | 0.973 | 3.476 | 3 | 0.681 |
| S_EMDAM_LSTM | 2.396 | 0.977 | 3.196 | 2 | 0.706 |
| **BiL-BERT** | **0.032** | **0.789** | **0.045** | **1** | **0.989** |

Table 24: The detailed results of the evaluation indicators for each model.

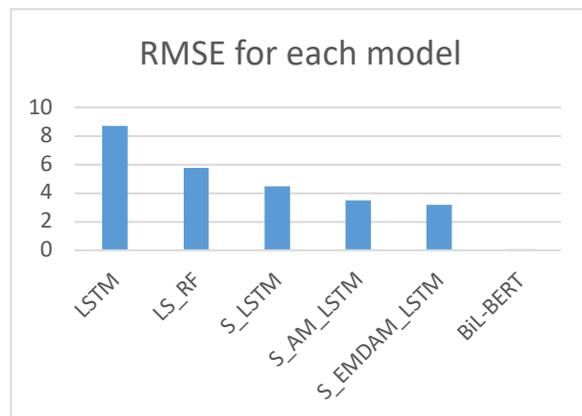

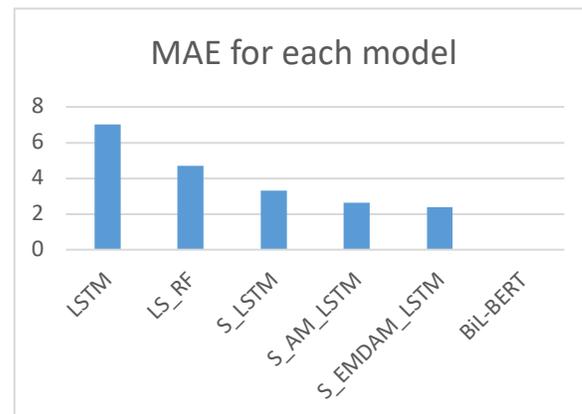

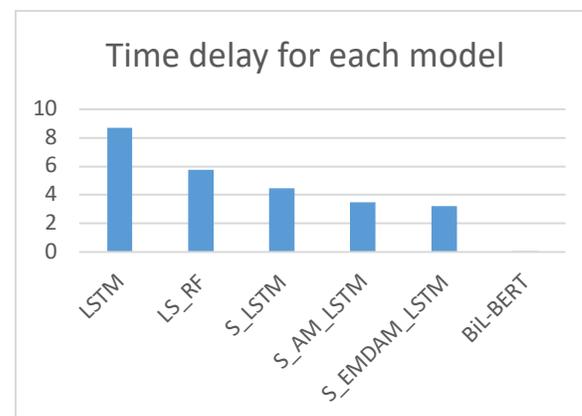

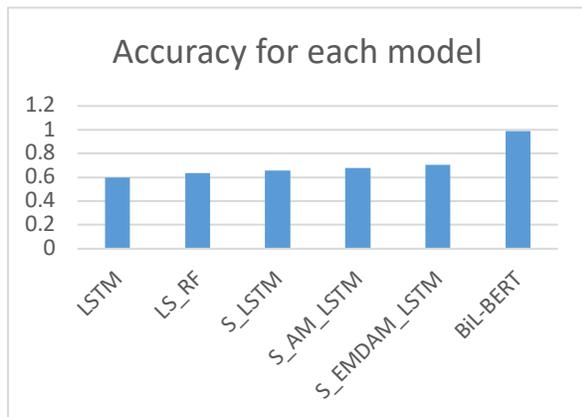

Figure 52, 53, 54, 55: The evaluation indicators for each model

## VII. FUTURE WORK

Though the other recent researchers have worked in single sector like Gold or only single index, in contrast with the proposed method which has been tested on almost all sectors of stocks, shares, and commodities in different stock exchanges, across different indices across the countries and has shown significant better results. Still the proposed model has some scopes for improvement. Here only the English tweets are considered for score calculation. Experiments on the effects of tweets and news posted in the languages other than English will be taken in future. Although the algorithm has well deterministic steps to measure the aggregate scores for multiple tweets (of different and opposite opinions) adding or nullifying the scores, future scope is there to include more tweets of opposite opinions and measure the final score.

## VIII. CONCLUSION

This is an original research work to predict the fluctuation of financial market based on tweets from popular and influential twitter handles. The proposed BilBERT model is first of its kind which can predict price of stock and share from only past data without twitter sentiment and considering the twitter sentiment combining with past stock data and able to show the difference. Most of the researchers have only worked in measuring the effect of some words on stock prices or have considered effect of tweets on only single share. The proposed BilBERT is able to consider cross and counter tweets for score calculation for predicting stock price of different sectors, different indices across the globe. It has outperformed all other available models considering features, flexibility, originality, and performance of prediction tested for stock and share price data of last 20 years due presented with results of tables and graphs.

## REFERENCES

1. Adam Atkins, Mahesan Niranjan, Enrico Gerding, "Financial news predicts stock market volatility better than close price" Volume 4, Issue 2, June 2018, Pages 120-137 ,https://doi.org/10.1016/j.jfds.2018.02.002
2. Xinyi Li , Yinchuan Li , Hongyang Yang , Liuqing Yang , Xiao-Yang Liu Columbia University, Beijing Institute of Technology, "DP-LSTM: Differential Privacy-inspired LSTM for Stock Prediction Using Financial News", arXiv preprint https://arXiv:1912.10806
3. Yinghao Ren, "2019 International Conference on Identification, Information and Knowledge in the Internet of Things (IIKI2019) Impact of News on the Trend of Stock Price", Procedia Computer Science 174 (2020) 128–140
4. Xin Du, "Stock Embeddings Acquired from News Articles and Price History, and an Application to Portfolio Optimization", 3353-3363. 10.18653/v1/2020.acl-main.307
5. Sunghyuck Hong, "A study on stock price prediction system based on text mining method using LSTM and stock market news", 223–28. doi:10.14400/JDC.2020.18.7.223
6. Isaac Kofi Nti, Adebayo Felix Adekoya , Benjamin Asubam Weyori, "Predicting Stock Market Price Movement Using Sentiment Analysis: Evidence From Ghana", Applied Computer Systems. 25. 33-42. 10.2478/acss-2020-0004.
7. Sandipan Biswas , Ahona Ghosh, Srabanti Chakraborty , Sandip Roy, Rajesh Bose, "Scope of Sentiment Analysis on News Articles Regarding Stock Market and GDP in Struggling Economic Condition", International Journal of Emerging Trends in Engineering Research. 8. 3594-3609. 10.30534/ijeter/2020/117872020.
8. Nur Ghaniaviyanto Ramadhan, Imelda Atastina, "Neural Network on Stock Prediction using the Stock Prices Feature and Indonesian Financial News Titles", https://doi.org/10.21108/ijoict.v7i1.544
9. Marah-Lisanne Thormann, Jan Farchmin, Christoph Weisser, Rene-Marcel Kruse , Benjamin Safken, Alexander Silbersdorff, "Stock Price Predictions with LSTM Neural Networks and Twitter Sentiment", Statistics, Optimization & Information Computing, 9(2), 268-287. https://doi.org/10.19139/soic-2310-5070-1202
10. Priyank Sonkiya, Vikas Bajpai, Anukriti Bansal, "Stock price prediction using BERT and GAN", arXiv preprint https://arxiv.org/abs/2107.09055
11. Zhenda Hu, "Crude oil price prediction using CEEMDAN and LSTM-attention with news sentiment index", Oil & Gas Science and Technology. 76. 28. 10.2516/ogst/2021010.
12. Mahtab Mohtasham Khani, Sahand Vahidnia, Alireza Abbasi, "A Deep Learning-Based Method for Forecasting Gold Price with Respect to Pandemics", SN COMPUT. SCI. 2, 335 (2021). https://doi.org/10.1007/s42979-021-00724-3

13. Petr Hajek, Josef Novotny, "Fuzzy Rule-Based Prediction of Gold Prices using News Affect", Expert Systems with Applications, Volume 193, 2022, 116487, ISSN 0957-4174, https://doi.org/10.1016/j.eswa.2021.116487.
14. Ye Ma, Lu Zong, Peiwan Wang, "A Novel Distributed Representation of News (DRNews) for Stock Market Predictions", arXiv preprint https://arxiv.org/abs/2005.11706
15. Taylan Kabbani, Fatih Enes Usta(2022), "Predicting The Stock Trend Using News Sentiment Analysis and Technical Indicators in Spark", arXiv preprint https://arxiv.org/abs/2201.12283
16. Ishu Gupta, Tarun Kumar Madan, Sukhman Singh, Ashutosh Kumar Singh, "HISA-SMFM: HISTORICAL AND SENTIMENT ANALYSIS BASED STOCK MARKET FORECASTING MODEL", arXiv preprint https://arxiv.org/abs/2203.08143
17. Shayan Halder, "FinBERT-LSTM: Deep Learning based stock price prediction using News Sentiment Analysis", arXiv preprint https://arxiv.org/abs/2211.07392
18. Zakaria Alameera,b , Mohamed Abd Elazizc , Ahmed A. Eweesd , Haiwang Yea, Zhang Jianhuaa, "Forecasting gold price fluctuations using improved multilayer perceptron neural network and whale optimization algorithm", Volume 61, 2019, Pages 250-260, ISSN 0301-4207, https://doi.org/10.1016/j.resourpol.2019.02.014.
19. Jessica, Raymond Sunardi Oetama, "Sentiment Analysis on Official News Accounts of Twitter Media in Predicting Facebook Stock", 2019 5th International Conference on New Media Studies (CONMEDIA), Bali, Indonesia, 2019, pp. 74-79, doi: 10.1109/CONMEDIA46929.2019.8981836.
20. Saloni Mohan, Sahitya Mullapudi, Sudheer Sammeta, Parag Vijayvergia and David C. Anastasiu, "Stock Price Prediction Using News Sentiment Analysis", 2019 IEEE Fifth International Conference on Big Data Computing Service and Applications (BigDataService), Newark, CA, USA, 2019, pp. 205-208, doi: 10.1109/BigDataService.2019.00035.
21. Yingzhe Dong, Da Yan, Abdullateef Ibrahim Almudaifer, Sibo Yan, Zhe Jiang, Yang Zhou, "BELT: A Pipeline for Stock Price Prediction Using News", 2020 IEEE International Conference on Big Data (Big Data), Atlanta, GA, USA, 2020, pp. 1137-1146, doi: 10.1109/BigData50022.2020.9378345.
22. Ioannis E. Livieris, Emmanuel Pintelas, Panagiotis Pintelas, "A CNN–LSTM model for gold price time-series forecasting", Neural Comput & Applic 32, 17351–17360 (2020). https://doi.org/10.1007/s00521-020-04867-x
23. Jingyi Shen and M. Omair Shafq, "Short-term stock market price trend prediction using a comprehensive deep learning system", J Big Data 7, 66 (2020). https://doi.org/10.1186/s40537-020-00333-6
24. Bipin Aasi, Syeda Aniqa Imtiaz, Hamzah Arif Qadeer, Magdalean Singarajah, Rasha Kashef, "Stock Price Prediction Using a Multivariate Multistep LSTM: A Sentiment and Public Engagement Analysis Model", 2021 IEEE International IOT, Electronics and Mechatronics Conference (IEMTRONICS), Toronto, ON, Canada, 2021, pp. 1-8, doi: 10.1109/IEMTRONICS52119.2021.9422526.
25. Wasiat Khan, Mustansar Ali Ghazanfar,Muhammad Awais Azam, Amin Karami, Khaled H. Alyoubi, Ahmed S. Alfakeeh, "Stock market prediction using machine learning classifers and social media, news", J Ambient Intell Human Comput 13, 3433–3456 (2022). https://doi.org/10.1007/s12652-020-01839-w
26. Naadun Sirimevan, I.G. U. H. Mamalgaha, Chandira Jayasekara, Y. S. Mayuran, and Chandimal Jayawardena, "Stock Market Prediction Using Machine Learning Techniques", Stock Market Prediction Using Machine Learning Techniques. 192-197. 10.1109/ICAC49085.2019.9103381.
27. Otabek Sattarov, Heung Seok Jeon, Ryumduck Oh and Jun Dong Lee, "Forecasting Bitcoin Price Fluctuation by Twitter Sentiment Analysis", 2020 International Conference on Information Science and Communications Technologies (ICISCT), Tashkent, Uzbekistan, 2020, pp. 1-4, doi: 10.1109/ICISCT50599.2020.9351527.
28. Padmanayana, Varsha, Bhavya K, "Stock Market Prediction Using Twitter Sentiment Analysis", International Journal of Scientific Research in Science and Technology. 265-270. 10.32628/CSEIT217475.
29. Ashwini Saini, Anoop Sharma, "Predicting the Unpredictable: An Application of Machine Learning Algorithms in Indian Stock Market", Ann. Data. Sci. 9, 791–799 (2022). https://doi.org/10.1007/s40745-019-00230-7
30. Jithin Eapen, Abhishek Verma, Doina Bein, "Novel Deep Learning Model with CNN and Bi-Directional LSTM for Improved Stock Market Index Prediction", 2019 IEEE 9th Annual Computing and Communication Workshop and Conference (CCWC), Las Vegas, NV, USA, 2019, pp. 0264-0270, doi: 10.1109/CCWC.2019.8666592.
31. Pengfei Yu, Xuesong Yan, "Stock price prediction based on deep neural networks", Neural Comput & Applic 32, 1609–1628 (2020). https://doi.org/10.1007/s00521-019-04212-x
32. Md. Arif Istiake Sunny, Mirza Mohd Shahriar Maswood, Abdullah G. Alharbi, "Deep Learning-Based Stock Price Prediction Using LSTM and Bi-Directional LSTM Model", 2020 2nd Novel Intelligent and Leading Emerging Sciences Conference (NILES), Giza, Egypt, 2020, pp. 87-92, doi: 10.1109/NILES50944.2020.9257950.

33. Sidra Mehtab, Jaydip Sen, "Stock Price Prediction Using CNN and LSTMBased Deep Learning Models", arXiv preprint https://arxiv.org/abs/2010.13891

34. Adil MOGHAR, Mhamed HAMICHE, "Stock Market Prediction Using LSTM Recurrent Neural Network", Procedia Computer Science, Volume 170, 2020, Pages 1168-1173, ISSN 1877-0509, https://doi.org/10.1016/j.procs.2020.03.049.

35. Irfan Ramzan Parray, Surinder Singh Khurana, Munish Kumar, Ali A. Altalbe, "Time series data analysis of stock price movement using machine learning techniques",

36. Zhigang Jin, Yang Yang, Yuhong Liu, "Stock closing price prediction based on sentiment analysis and LSTM",Neural Computing and Applications(2019). doi: 10.1007/s00521-019-04504-2